\documentstyle[twoside]{article}

\catcode`\@=11
\long\def\@makefntext#1{ %\parindent 1em
\protect\noindent \hbox to 3.2pt {\hskip-.9pt
$^{{\eightrm\@thefnmark}}$\hfil}#1\hfill} %can be used

\def\thefootnote{\fnsymbol{footnote}}
 \def\@makefnmark{\hbox to 0pt{$^{\@thefnmark}$\hss}}  %original

\def\ps@myheadings{\let\@mkboth\@gobbletwo
\def\@oddhead{\hbox{} %\sl
\rightmark\hfil\eightrm\thepage}
\def\@oddfoot{}\def\@evenhead{\eightrm\thepage\hfil %\sl
\leftmark\hbox{}}\def\@evenfoot{}
\def\sectionmark##1{}\def\subsectionmark##1{}}

\oddsidemargin=\evensidemargin
\renewcommand{\thefootnote}{\fnsymbol{footnote}}

\newcounter{sectionc}\newcounter{subsectionc}\newcounter{subsubsectionc}
\renewcommand{\section}[1] {\vspace{12pt}\addtocounter{sectionc}{1}
\setcounter{subsectionc}{0}\setcounter{subsubsectionc}{0}\noindent
	{\tenbf\thesectionc. #1}\par\vspace{5pt}}
\renewcommand{\subsection}[1] {\vspace{12pt}\addtocounter{subsectionc}{1}
	\setcounter{subsubsectionc}{0}\noindent
	{\bf\thesectionc.\thesubsectionc. {\kern1pt \bfit #1}}\par\vspace{5pt}}
\renewcommand{\subsubsection}[1] {\vspace{12pt}\addtocounter{subsubsectionc}{1}
	\noindent{\tenrm\thesectionc.\thesubsectionc.\thesubsubsectionc.
	{\kern1pt \tenit #1}}\par\vspace{5pt}}
\newcommand{\nonumsection}[1] {\vspace{12pt}\noindent{\tenbf #1}
	\par\vspace{5pt}}

\newcounter{appendixc}
\newcounter{subappendixc}[appendixc]
\newcounter{subsubappendixc}[subappendixc]
\renewcommand{\thesubappendixc}{\Alph{appendixc}.\arabic{subappendixc}}
\renewcommand{\thesubsubappendixc}
	{\Alph{appendixc}.\arabic{subappendixc}.\arabic{subsubappendixc}}

\renewcommand{\appendix}[1] {\vspace{12pt}
        \refstepcounter{appendixc}
        \setcounter{figure}{0}
        \setcounter{table}{0}
        \setcounter{lemma}{0}
        \setcounter{theorem}{0}
        \setcounter{corollary}{0}
        \setcounter{definition}{0}
        \setcounter{equation}{0}
        \renewcommand{\thefigure}{\Alph{appendixc}.\arabic{figure}}
        \renewcommand{\thetable}{\Alph{appendixc}.\arabic{table}}
        \renewcommand{\theappendixc}{\Alph{appendixc}}
        \renewcommand{\thelemma}{\Alph{appendixc}.\arabic{lemma}}
        \renewcommand{\thetheorem}{\Alph{appendixc}.\arabic{theorem}}
        \renewcommand{\thedefinition}{\Alph{appendixc}.\arabic{definition}}
        \renewcommand{\thecorollary}{\Alph{appendixc}.\arabic{corollary}}
        \renewcommand{\theequation}{\Alph{appendixc}.\arabic{equation}}
        \noindent{\tenbf Appendix \theappendixc #1}\par\vspace{5pt}}
\newcommand{\subappendix}[1] {\vspace{12pt}
        \refstepcounter{subappendixc}
        \noindent{\bf Appendix \thesubappendixc. {\kern1pt \bfit #1}}
	\par\vspace{5pt}}
\newcommand{\subsubappendix}[1] {\vspace{12pt}
        \refstepcounter{subsubappendixc}
        \noindent{\rm Appendix \thesubsubappendixc. {\kern1pt \tenit #1}}
	\par\vspace{5pt}}

\newcommand{\textlineskip}{\baselineskip=13pt}
\newcommand{\smalllineskip}{\baselineskip=10pt}

\newcommand{\copyrightheading}[1]
	{\hspace*{9cm}\smalllineskip{\flushright
	{\tenrm ULB-PMIF/93-06}}}
\def\abstracts#1#2#3{{
	\centering{\begin{minipage}{4.5in}\baselineskip=10pt\eightrm
	\centerline{ABSTRACT}
	\parindent=0pt #1\par
	\parindent=15pt #2\par
	\parindent=15pt #3
	\end{minipage} }\par}}

\newcommand{\bibit}{\nineit}

\renewenvironment{thebibliography}[1]			%ALL CHANGES DD 13/3/92
	{\ninerm
	 \baselineskip=11pt				%changed by cheng
	 \begin{list}{\arabic{enumi}.}
	{\usecounter{enumi}\setlength{\parsep}{0pt}
	 \setlength{\leftmargin 17pt}{\rightmargin 0pt}	%changed by cheng
	 \setlength{\itemsep}{0pt} \settowidth		%changed by cheng
	{\labelwidth}{#1.}\sloppy}}{\end{list}}

\newcounter{itemlistc}
\newcounter{romanlistc}
\newcounter{alphlistc}
\newcounter{arabiclistc}

\newcommand{\fcaption}[1]{
        \refstepcounter{figure}
        \setbox\@tempboxa = \hbox{\eightrm Fig.~\thefigure. #1}
        \ifdim \wd\@tempboxa > 5in
           {\begin{center}
        \parbox{5in}{\eightrm \smalllineskip Fig.~\thefigure. #1 }
            \end{center}}
        \else
             {\begin{center}
             {\eightrm Fig.~\thefigure. #1}
              \end{center}}
        \fi}

\newcommand{\tcaption}[1]{
        \refstepcounter{table}
        \setbox\@tempboxa = \hbox{\eightrm Table~\thetable. #1}
        \ifdim \wd\@tempboxa > 5in
           {\begin{center}
        \parbox{5in}{\eightrm\smalllineskip Table~\thetable. #1 }
            \end{center}}
        \else
             {\begin{center}
             {\eightrm Table~\thetable. #1}
              \end{center}}
        \fi}

\def\@citex[#1]#2{\if@filesw\immediate\write\@auxout	%IJMPA, IJMPB ONLY
	{\string\citation{#2}}\fi			%TO DELETE PERCENTAGE
\def\@citea{}\@cite{\@for\@citeb:=#2\do			%KEY WHEN USING
	{\@citea\def\@citea{,}\@ifundefined		%DD 13/3/92
	{b@\@citeb}{{\bf ?}\@warning
	{Citation `\@citeb' on page \thepage \space undefined}}
	{\csname b@\@citeb\endcsname}}}{#1}}

\newif\if@cghi
\def\cite{\@cghitrue\@ifnextchar [{\@tempswatrue
	\@citex}{\@tempswafalse\@citex[]}}
\def\citelow{\@cghifalse\@ifnextchar [{\@tempswatrue
	\@citex}{\@tempswafalse\@citex[]}}
\def\@cite#1#2{{$\null^{#1}$\if@tempswa\typeout
	{IJCGA warning: optional citation argument
	ignored: `#2'} \fi}}

\def\pmb#1{\setbox0=\hbox{#1}
	\kern-.025em\copy0\kern-\wd0
	\kern.05em\copy0\kern-\wd0
	\kern-.025em\raise.0433em\box0}

\def\fnt#1#2{\footnotetext{\kern-.3em
	{$^{\mbox{\scriptsize #1}}$}{#2}}}

\def\fpage#1{\begingroup
\voffset=.3in
\thispagestyle{empty}\begin{table}[b]\centerline{\footnotesize #1}
	\end{table}\endgroup}

\def\runninghead#1#2{\pagestyle{myheadings}
\markboth{{\eightit{\quad #1}}\hfill}{\hfill{\eightit{#2\quad}}}}

\font\tenbf=cmbx10
\font\tenit=cmti10
\font\tenit=cmti10
\font\bfit=cmbxti10 at 10pt
 1
 1
 1

\font\ninerm=cmr9
\font\nineit=cmti9

\font\eightrm=cmr8
\font\eightit=cmti8

\def\qed{\hbox{${\vcenter{\vbox{                          %HOLLOW SQUARE
   \hrule height 0.4pt\hbox{\vrule width 0.4pt height 6pt
   \kern5pt\vrule width 0.4pt}\hrule height 0.4pt}}}$}}

\runninghead{Anomalies in the Hamiltonian Formalism.
} {A Cohomological Analysis.
}

\begin{document}
\normalsize\textlineskip
{\thispagestyle{empty}
\setcounter{page}{1}

\renewcommand{\thefootnote}{\fnsymbol{footnote}} %use symbolic footnote

%print out the publisher copyright heading
\copyrightheading{}

\vspace*{0.88truein}

\fpage{1}
\centerline{\bf PERTURBATIVE GAUGE ANOMALIES}
\vspace*{0.035truein}
\centerline{\bf IN THE HAMILTONIAN FORMALISM:}
\vspace*{0.035truein}
\centerline{\bf A COHOMOLOGICAL ANALYSIS.}
\vspace{0.37truein}
\centerline{\footnotesize G. BARNICH\footnote{Aspirant au Fonds
National de la Recherche Scientifique (Belgium).}}
\vspace*{0.015truein}
\centerline{\footnotesize\it Facult\'e des Sciences,
Universit\'e Libre de Bruxelles,}
\baselineskip=10pt
\centerline{\footnotesize\it Campus Plaine C.P. 231, B-1050 Bruxelles,
Belgium}
\vspace{0.225truein}

\abstracts{\noindent The quantum action principle of renormalisation theory
is applied to the antibracket-antifield formalism for Hamiltonian systems.
General results on the local BRST cohomology allow one to prove that the
anomalies appear in the time development of the BRST charge and violate
the nilpotency of this charge.
Furthermore they are equivalent to those of the Lagrangian formalism.
The analysis provides a completely gauge and
regularisation independent proof of Faddeev's conjecture on the relationship
between gauge anomalies and Schwinger terms in the context of descent
equations.}{}{}
\vspace*{-3pt}\textlineskip
\textheight=7.8truein
\setcounter{footnote}{0}
\renewcommand{\thefootnote}{\alph{footnote}}
\vspace{0.37truein}
\section{Introduction}
\noindent
The existence of a relationship between gauge anomalies and the appearance
of Schwinger terms in the equal time commutation relations between the
corresponding currents was first established by perturbative
calculations\cite{Adler,Jackiw}.
The discovery that anomalies are constrained by consistency
conditions\cite{Wess} and the use of the quantum action
principle\cite{Lowenstein,Lam} allowed to determine the existence of
anomalies and to calculate their form by cohomological techniques\cite{Becchi}.
It is natural that there should exist a related algebraic principle
constraining the existence of possible Schwinger terms. For non-abelian
anomalies, the first investigation in this direction\cite{Faddeev} in the
Hamiltonian formalism showed that the Schwinger terms should be related
to the gauge anomaly through descent equations\cite{Stora,Zumino}.
It was followed by a series of perturbative calculations\cite{Jo} in order
to verify this conjecture.

Recently\cite{Fujiwara} Hamiltonian BRST methods\cite{Fradkin,Henneaux}
have been used and an algebraic principle based on locality assumptions and
the validity of the Jacobi identity for the equal time commutator involving
the BRST charge and the Hamiltonian has been conjectured.
The authors of\cite{Fujiwara} then infer that the anomaly
in $[\hat\Omega,\hat H]$ corresponds to the
Lagrangian anomaly and that the link with the the anomaly in
$[\hat\Omega,\hat\Omega]$ should be
understood by some some sort of descent. As they are careful to point out,
they are interested above all in the calculational aspect of anomalies and
their results are not meant to be rigorous because no discussion of
renormalisation in the Hamiltonian formalism is made.

The purpose of this letter is the rigorous treatement of anomalies in the
Hamiltonian framework.
We will be able to prove the basic equations about Hamiltonian anomalies
conjectured in\cite{Fujiwara} (see also\cite{Baulieu}).
The main idea is to quantize perturbatively our first order Hamiltonian
system like a Lagrangian one. It has already been
shown\cite{Henneaux,Dresse} that this gives formally equivalent results to
those obtained from a perturbative quantization of the underlying
Lagrangian system. There is no need for a new algebraic
consistency condition because the quantum action principle applies and, like
in the Lagrangian case, it can be used to reduce the whole analysis of the
first order renormalisation effects to a local cohomological problem, which
amounts to finding the general solution to a set of descent equations.

We show that, by choosing appropriate representatives in the relevant
cohomological classes, the content of these descent equations can be
expressed in terms of quantities of the Hamiltonian formalism.
In terms of these representatives, the relationship
through descent equations of the anomaly
in the time development of the BRST charge and the violation of its
nilpotency naturally follows (Faddeev's conjecture).
Furthermore, the equivalence to the Lagrangian anomalies is also direct,
because of a theorem\cite{Barnich} proving the invariance of local BRST
cohomology classes with respect to generalized auxiliary fields content,
which is precisely what the Langrangian and the Hamiltonian
antibracket-antifield formalisms differ in.

Regularisation independence of the analysis is guaranteed by using general
results from renormalisation theory, like the quantum action principle,
while gauge independence holds for all gauges of the antibracket-antifield
formalism.

\section{Quantum action principle and antibracket-antifield formalism.}
\noindent
Let us briefly recall the application of the quantum action principle
(see\cite{Piguet} for a review) in the context of the antibracket-antifield
formalism\cite{Howe}\footnote{For a different point of view on anomalies in
this formalism see also\cite{Troost}.}.
{}From a classical gauge theory with a local action $S_0[\phi^i]$, one builts
a possibly non-minimal proper and local solution of the
classical\footnote{It is sufficient
for the first order quantum corrections to consider
the classical master equation, because the corrections of the quantum master
equation are ill-defined and drop out anyway in
the course of (BPHZ) renormalisation (see\cite{Lam1}).}$\ $ master
equation\cite{Batalin1,Henneaux} $(S,S)=0$.
A gauge fixed action is obtained by making a canonical (in the antibracket
sense) change of variables which consists either of shifting the antifields
with the help of some local gauge fixing fermion $\Psi$ or of exchanging
for some field-antifield pair the role of the field and the antifield
(including a minus sign) $(\phi^A,\phi_A^*)\longrightarrow
(-\phi_A^*, \phi^A)$ in such a way that after putting to zero all the (new)
antifields, the action has no gauge freedom left.
One then considers the extended action
$S_{ext}[\phi^A;\phi_A^*]=S[\phi^A;\phi_A^* + \delta\Psi /\delta\phi^A]$.
The Legendre transform on the sources $J_A$ of the generating functional
$Z_c[J_A;\phi_A^*]$ for connected Green's functions associated to
$S_{ext}[\phi^A;\phi_A^*]$ gives the generating functional for one particle
irreducible Green's functions $\Gamma[\phi^A_{cl};\phi_A^*]$.
At tree level, $\Gamma[\phi^A_{cl};\phi_A^*]$ is equal to
$S_{ext}[\phi^A_{cl};\phi_A^*]$ which satisfies the master equation
$(S_{ext},S_{ext})$ in the sources $\phi^A_{cl}$ and $\phi_A^*$.
The quantum action principle then states that\cite{Zinn-Justin}
\begin{equation}
(\Gamma,\Gamma)=[\Delta\cdot\Gamma]\label{eq1}
\end{equation}
where $\Delta$ is a local integrated polynomial
of ghost number $1$, of fixed dimension and of order at least one in $\hbar$.
Here $[\Delta\cdot\Gamma]$ denotes the generating functional for one
particle irreducible Green's functions with the insertion $\Delta$.
The quantum action principle is a general result from renormalisation theory;
it has been proved in various renormalisation schemes and is believed to be
scheme independent.
At lowest order the identity $((\Gamma,\Gamma),\Gamma)=0$ yields the
consistency condition $(\Delta,S_{ext})=0$ in the sources $\phi^A_{cl}$ and
$\phi_A^*$ on the insertion $\Delta$.
Solutions of the form $\Delta=(\Lambda,S_{ext})$, with $\Lambda$ a local
integrated polynomial of ghost number zero and the dimension of $S_{ext}$,
can be absorbed through local counterterms added to $S_{ext}$.

Dropping the subscript in  $\phi^A_{cl}$ and making the canonical change of
variables $\phi_A^* \longrightarrow \phi_A^* - \delta\Psi/\delta\phi^A$
we have to find all cohomologically non trivial solutions of
\begin{equation}
(\Delta^\prime,S) =0\label{consistencycond}
\end{equation}
with $\Delta^\prime[\phi^A,\phi_A^*] = \Delta[\phi^A,\phi_A^* -
\delta\Psi/\delta\phi^A]$.
Let $\Delta^\prime = \int b_{[k]}$ where $b_{[k]}$ is a $D-$form valued
polynomial in the fields, the antifields and a finite number of their
derivatives.
The boundary conditions on the fields and antifields (they are sources and
as such they are $C^\infty$ fast decreasing functions), imply that
(\ref{consistencycond}) is equivalent to (see for instance\cite{Henneaux}
chapter 12)
\begin{equation}
s b_{[k]} + d b_{[k-1]} =0 \label{localconsist}.
\end{equation}
A non trivial solution $b_{[k]}$ is an element of $H(s|d)$ in form degree
$D$ whose shortest descent stops after $k$ steps. Indeed, because of the
triviality\cite{Dubois-Violette} of the cohomology of the algebraic exterior
differential $d$, (\ref{localconsist}) yields a set of descent equations
\begin{eqnarray}
s b_{[k-1]} + d b_{[k-2]} =0\\
\vdots\nonumber\\
s b_{[0]}=0 \label{bottom}.
\end{eqnarray}
The strategy\cite{Dubois-Violette1} to find the most general non-trivial
solution of (\ref{localconsist}) is the following: first one looks for the
most general solution of the bottom equation (\ref{bottom}), i.e., one
chooses all $s$ mod $d$ nontrivial solutions of
$H(s)$ in all form degrees\footnote{If the bottom is $s$ mod $d$ trivial, a
redefinition of $b_{[k]}$ allows to get a shorter set of descent equations.}.
Then one tries if
the solutions $b_{[0]}$ in form degree $D-k$
can be lifted $k$ steps to yield a non-trivial solution of
(\ref{localconsist}).

Because the usual non-minimal sectors introduced for gauge fixing purposes
do not contain derivatives of the additional fields, the contracting
homotopy, which eliminates these fields from $H(s)$, commutes with
derivatives (see\cite{Henneaux} chapter 12) and they do not appear in
$H(s|d)$ either, implying that it is enough to consider $s$ associated to a
minimal solution of the master equation.

\section{Application to Hamitonian systems}
\noindent
\subsection{Descent equations in the Hamiltonian framework}
\noindent
Consider a local extended Hamiltonian formalism satisfying suitable
regularity

\noindent conditions.
The BRST charge $\Omega$ and the Hamiltonian $H$ can be constructed as
local functionals in space,
$\Omega =\int d^{D-1}x\  \omega,\ H=\int d^{D-1}x\  h$, where $\omega$ and
$h$ depend on the fields $\{q^i(t,x^m),p_i(t,x^m),\eta^{a}(t,x^m),
{\cal P}_{a}(t,x^m)\}\equiv \phi^A (t,x^m)$,
$m=1,...,D-1$ and a finite number of their spatial derivatives $\partial_m$.
A local proper solution to the master equation is
given by\cite{Henneaux,Dresse}
\begin{eqnarray}
S_H[\phi^A,\phi_A^*]=\int d^Dx {\cal L}_H=\int dt\int d^{D-1}x -
{1\over 2}\dot\phi^A \sigma_{AB}\phi^B - h \nonumber\\
+{{\stackrel{\leftarrow}{\partial}\omega\over{\partial (\phi^A})^{0(k)}}}
\sigma^{AB}(\phi_B^*)^{0(k)},
\end{eqnarray}
where we have made the identification ${\cal P}_a=-\lambda^*_a$, with
$\lambda^a$ the Lagrange multipliers for the constraints\footnote{In the
index $l(k)$, $l$ refers to the time while $(k)$ is a space multiindex.}.
For notational simplicity we consider only irreducible constraints, the
reducible case can be analyzed along the same lines\cite{Henneaux}.
If one puts the antifields to zero, the gauge is
completely fixed (multiplier gauge $\lambda^a = 0$).
The local nilpotent BRST symmetry\footnote{${\tilde\delta}/\tilde\delta$ is
the Euler-Lagrange derivative in space.} $s_H$:
\begin{eqnarray}
s_H  = {\stackrel{\leftarrow}{\partial}\over \partial (\phi^A)^{l(k)}}
(\sigma^{AB}{\stackrel{\rightarrow}{\tilde\delta}
\over\tilde\delta \phi^B}\omega)^{l(k)}\nonumber\\
-{\stackrel{\leftarrow}{\partial}\over {\partial (\phi_A^*})^{l(k)}}
(\sigma_{AB}\dot\phi^B-{\stackrel{\rightarrow}{\tilde\delta h}\over
\tilde\delta \phi^A}
+{\stackrel{\rightarrow}{\tilde\delta}\over \tilde\delta \phi^A}
{{\stackrel{\leftarrow}{\partial}\omega\over {\partial (\phi^A})^{0(m)}}}
\sigma^{AB}(\phi_B^*)^{0(m)})^{l(k)}.
\end{eqnarray}
splits according to the antifield number into the sum of $\delta$, the
Koszul-Tate differential of the
Hamiltonian stationary surface, and $\sigma = s_\omega + \gamma$, where
\begin{eqnarray}
s_\omega = {{\stackrel{\leftarrow}{\partial}
\over {\partial (\phi^A})^{0(k)}}}
(\sigma^{AB}{\stackrel{\rightarrow}{\tilde\delta}
\over\tilde\delta \phi^B}\omega)^{0(k)}
\end{eqnarray}
is the local version of the usual Hamiltonian BRST symmetry.
Take
\begin{eqnarray}
\rho = -{{\stackrel{\leftarrow}{\partial}\over {\partial (\phi^A})^{l+1(k)}}}
\sigma^{AB}(\phi_B^*)^{l(k)}.
\end{eqnarray}
Consider a representative of an element of $H(s_H)$, $b\ \in {\cal {FA}}$,
where ${\cal {FA}}$ is the space of form valued polynomials in the fields,
the antifields and a finite number of their derivatives,
and apply the anticommutator of $\rho$  with $s_H$ to homogeneous terms in
$b$ containing $n$ antifields and $m$ fields with at least one time
derivative, $n+m\neq 0$, to get
\begin{eqnarray}
(n+m)b - s_H \rho b=
-{\stackrel{\leftarrow}{\partial} b \over {\partial (\phi^A})^{l+1(k)}}
\sigma^{AB}
(-{\stackrel{\rightarrow}{\tilde\delta h}\over \tilde\delta \phi^A}
+{\stackrel{\rightarrow}{\tilde\delta}\over \tilde\delta \phi^A}
{{\stackrel{\leftarrow}{\partial}\omega\over {\partial (\phi^A})^{0(m)}}}
\sigma^{AB}(\phi_B^*)^{0(m)})^{l(k)}\label{fund}.
\end{eqnarray}
This means that, by repeated redefinitions, one can first absorb all
time derivatives of the fields and then the antifields.
A representative of an element of $H(s_H)$ can be chosen
from $\tilde {\cal F}$, the space of form valued polynomials in the fields
and a finite number of their spatial derivatives.
Because on such representatives, the cocycle condition $s_Hb=0$ reduces to
$s_\omega b=0$, and the freedom of adding $s_H$-exact terms is reduced to
$s_\omega$-exact terms we get that
$H(s_H)$ in ${\cal {FA}}$ is isomorphic to $H(s_\omega)$ in
$\tilde {\cal F}$.

Let $d=\tilde d + d^0$ where $\tilde d = dx^m\partial_m$ and
$d^0=dt\partial_t$.
The bottom $b_{[0]}\ \in \tilde {\cal F}$ of a set descent equations has to
be a $s_\omega$ mod $\tilde d$ non-trivial element of  $H(s_\omega)$.
Trying to lift $b_{[0]}$, we have to solve the equation
\begin{equation}
s_Hb_{[1]}+db_{[0]}=0\label{anti1}.
\end{equation}
Applying $\rho s_H +s_H\rho$ to $b_{[1]}$, we get (\ref{fund}) with the
additional term $-\rho d b_{[0]}$ on the right hand side.
Because this term contains no time
derivatives of the fields, $b_{[1]}$ can be chosen to be independent of the
time derivatives of the fields as well. The acyclicity
of $\delta$ in ${\cal {FA}}$, then implies that
all terms with antifield number higher than $2$ can be absorbed.
If $b_{[1]}=dtb_{[1]}^0 +\tilde b_{[1]}$ where
$b_{[1]}^0 ,\tilde b_{[1]}$ are independent of $dt$,
(\ref{anti1}) splits according to antifield number and $dt$ into
\begin{eqnarray}
\delta b_{[1]1}^0 + s_\omega b_{[1]0}^0+\partial_t\tilde b_{[0]} -
\tilde d b_{[0]}=0\\
\delta\tilde b_{[1]1} +s_\omega\tilde b_{[1]0}+\tilde d \tilde b_{[0]}=0\\
\sigma\tilde b_{[1]1}=0\ \ \sigma b_{[1]1}^0=0.
\end{eqnarray}
A necessary condition is that the terms multiplying the time derivatives of
the fields agree on both sides of these equations.
This implies that
\begin{equation}
b_{[1]1}^0 =
{\stackrel{\leftarrow}{\partial}\tilde b_{[0]}\over\partial
(\phi^A)^{0(k)}}\sigma^{AB}(\phi^{*}_{B})^{0(k)}
\end{equation}
and $\delta\tilde b_{[1]1}=0$ meaning that $\tilde b_{[1]1}$ can be absorbed
through redefinitions of
$b_{[1]}$.
We then get equations involving elements of $\tilde {\cal F}$ alone:
\begin{eqnarray}
s_\omega b_{[1]0}^0+{\stackrel{\leftarrow}{\partial}\tilde b_{[0]}
\over\partial (\phi^A)^{0(k)}}\sigma^{AB}
({\stackrel{\rightarrow}{{\tilde\delta}}h\over\tilde\delta\phi^B})^{0(k)}
- \tilde d b_{[0]}^0=0\label{f1}\\
s_\omega\tilde b_{[1]0}+\tilde d \tilde b_{[0]}=0 \label{f2}
\end{eqnarray}
Furthermore, $\sigma b_{[1]1}^0=0$ can be shown to be satisfied because
$s_\omega \tilde b_{[0]}=0$.

Trying to lift $b_{[1]}$ in the same way,
we find that each $b_{[l]}$, $0\leq l\leq k$ can be choosen independent of
the time derivatives of the fields and at most linear
in the antifields (acting with $d$, no time derivative acts on the antifield
dependent part of $b_{[l-1]}$), with
\begin{equation}
b_{[l]}=
dt ({\stackrel{\leftarrow}{\partial}\tilde b_{[l-1]0}\over\partial
(\phi^A)^{0(k)}}\sigma^{AB}(\phi^{*}_{B})^{0(k)}) + dt b_{[l]0}^0 +
\tilde b_{[l]0}\label{anom}
\end{equation}
where $\tilde b_{[k]0}=0$ and verifying
\begin{eqnarray}
s_\omega b_{[l]0}^0+{\stackrel{\leftarrow}{\partial}\tilde b_{[l-1]0}
\over\partial (\phi^A)^{0(k)}}\sigma^{AB}
({\stackrel{\rightarrow}{{\tilde\delta}}h\over\tilde\delta\phi^B})^{0(k)}
- \tilde d b_{[l-1]0}^0=0\label{f3}\\
s_\omega\tilde b_{[l]0}+\tilde d \tilde b_{[l-1]0}=0. \label{f4}
\end{eqnarray}
The equation
\begin{equation}
\sigma {\stackrel{\leftarrow}{\partial}\tilde b_{[l-1]0}\over\partial
(\phi^A)^{0(k)}}\sigma^{AB}(\phi^{*}_{B})^{0(k)}= \tilde d
{\stackrel{\leftarrow}{\partial}\tilde b_{[l-2]0}\over\partial
(\phi^A)^{0(k)}}\sigma^{AB}(\phi^{*}_{B})^{0(k)}
\end{equation}
is again satisfied because $s_\omega\tilde b_{[l-1]0}+\tilde d
\tilde b_{[l-2]0}=0$.

Hence solving the descent equation for $s_H$ is equivalent to finding first
the most general solution of the
(spatial) descent equations associated to $s_\omega$,
in maximal spatial form degree $D-1$, and then selecting those for which
there exist solutions
$b_{[k]}^0$ and $b_{[k-1]}^0$ satisfying (\ref{f3}).

\subsection{Identification of the anomalies}
\noindent
Let us analyze in more detail the left hand side of (\ref{eq1}) to identify
the classical relations which acquire quantum corrections:
\begin{eqnarray}
{\stackrel{\leftarrow}{\delta}\over\delta\phi^A}{\cal L}_H
{\stackrel{\rightarrow}{\delta}\over\delta\phi^*_A}{\cal L}_H=
-\dot\omega+\partial_k
((\dot\phi^A)^{0(l)}{\stackrel{\rightarrow}{\delta}\omega\over
{\delta (\phi^A})^{0(l)+e_k}})
-[h,\omega]_{loc}\nonumber\\
-{1\over 2}{\stackrel{\leftarrow}{\partial}\over (\phi^A)^{0(l)}}
[\omega,\omega]_{loc}\sigma^{AB}(\phi^*_B)^{0(l)}\label{eq2}.
\end{eqnarray}
Here $[\cdot,\cdot]_{loc}$ is the extended local Poisson bracket defined in
terms of the spatial Euler-Lagrange derivatives.
This means\cite{Piguet} that after integration over space and putting the
antifields to zero
\begin{equation}
{d\over dt} \Omega_{op} = {1\over 2}\hbar \int d^{d-1}x\ dt
(b_{[k]0}^0)_{op}+O(\hbar^2).
\end{equation}

On the other hand, writing explicitely the Ward identities associated
to (\ref{eq1}), we get\cite{Lowenstein1}
\begin{eqnarray}
\int d^D x\ < T N_\rho [{\stackrel{\rightarrow}{\delta}{\cal L}_H
\over\delta\phi^*_A}(x)]
{\stackrel{\rightarrow}{\delta}\over\delta\phi^A(x)}
\prod_j\phi^{A_j}(y_j) >\nonumber\\
= i \int d^D x\ < T N_{\rho +4-d_\alpha}
[{\stackrel{\leftarrow}{\delta}\over\delta\phi^A}{\cal L}_H
{\stackrel{\rightarrow}{\delta}\over\delta\phi^*_A}{\cal L}_H(x)+
{1\over 2}\hbar b_{[k]}(x) +O(\hbar^2)]\prod_j\phi^{A_j}(y_j) >.
\end{eqnarray}
Using (\ref{anom},\ref{eq2}), we find, identifying the terms according to
the antifields, that the classical relations
\begin{eqnarray}
[H,\Omega]=0,\ \ [\Omega,\Omega]=0,
\end{eqnarray}
where the extended Poisson bracket $[\cdot,\cdot]$ is defined with
functional derivatives in space for local integrated functionals,
get modified inside Green's functions by the following
quantum corrections\footnote{The Jacobi identity for $[\cdot,\cdot]_Q$
involving the BRST charge alone and the Hamiltonian with two copies of the
BRST charge is related to the identity $(\Gamma_H,(\Gamma_H,\Gamma_H))=0$, but
the discussion involves properties of the normal product operator insertions
of the BPHZ renormalisation scheme going beyond the scope of this letter.}
\begin{eqnarray}
\int d^D x [H,\Omega]_Q (x) =  {1\over 2}\hbar \int dt b^0_{[k]0}
+O(\hbar^2)\\
\int d^D x [\Omega,\Omega]_Q (x) = \hbar \int dt \tilde b_{[k-1]0}
+O(\hbar^2).
\end{eqnarray}

\subsection{Equivalence with Lagrangian anomalies}
\noindent
The Lagrangian formalism and the total Hamiltonian formalism are related by
auxiliary fields, i.e., the momenta and the Lagrange multipliers can be
eliminated from the total Hamiltonian action in an algebraic way by
means of their equations of motion to yield the Lagrangian action.
For the passage to the extended Hamiltonian formalism, a generalization of
the concept of auxiliary fields on the level of the solution of the master
equation has to be used. This is possible under general regularity
conditions, excluding in particular systems which
do not satisfy Dirac's conjecture\cite{Henneaux,Dresse}.
If the passage from the Lagrangian to the extended Hamiltonian formalism
can be done by means of generalized auxiliary fields preserving the
locality of both formalisms, then the following theorem
proved in\cite{Barnich} guarantees the equivalence of the first order
anomalous corrections in both formalisms:

The local cohomology classes $H(s|d)$ are isomorphic for theories differing
in generalized auxiliary field content.

\section{Conclusion}
\noindent

The use of the antibracket-antifield formalism and its cohomological
properties allows to show the equivalence of the first order anomalies of
the Lagrangian and the Hamiltonian formalism and to
prove, in a completely gauge and regularisation independent way, that
the anomalies in the Hamiltonian framework occur only in the time
development of the BRST charge and in the violation of
nilpotency of this charge. Within the descent
equations adapted to the Hamiltonian formalism, the relationship of both
anomalies, the gauge anomaly and the Schwinger terms in the Poisson bracket
algebra of the constraints naturally follows.

\nonumsection{Acknowledgements}
\noindent
The author is grateful to Marc Henneaux for suggesting the problem and to
Olivier Coussaert for useful discussions.

\nonumsection{References}

\end{document}